\documentclass{llncs} %

\usepackage{float}

\usepackage{booktabs}
\usepackage{graphicx}
\usepackage{wrapfig}
\usepackage{comment}
\usepackage[table]{xcolor}

\usepackage[labelfont=bf]{caption}
\usepackage[colorlinks=true,linkcolor=blue,citecolor=blue]{hyperref}

\usepackage{color,soul}
\usepackage{amsmath} 
\usepackage{mathtools} 
\usepackage{amsfonts} 
\usepackage{amssymb}

\usepackage{array}
\usepackage{rotating}
\usepackage{multirow}
\usepackage{multicol}
\usepackage{lscape}
\usepackage{subfig}

\usepackage[export]{adjustbox}

\usepackage[ruled,vlined]{algorithm2e}
\usepackage{xargs}                      

\usepackage[colorinlistoftodos,prependcaption,textsize=tiny]{todonotes}
\usepackage{multirow,tabularx}

\renewcommand{\thetable}{\arabic{table}}

\usepackage{color}
\definecolor{reddish}{rgb}{0.82, 0.1, 0.26}

\usepackage{tikz,xcolor,hyperref}

\definecolor{lime}{HTML}{A6CE39}
\DeclareRobustCommand{\orcidicon}{
	\begin{tikzpicture}
	\draw[lime, fill=lime] (0,0) 
	circle [radius=0.16] 
	node[white] {{\fontfamily{qag}\selectfont \tiny ID}};
	\draw[white, fill=white] (-0.0625,0.095) 
	circle [radius=0.007];
	\end{tikzpicture}
	\hspace{-2mm}
}

\foreach \x in {A, ..., Z}{\expandafter\xdef\csname orcid\x\endcsname{\noexpand\href{https://orcid.org/\csname orcidauthor\x\endcsname}
			{\noexpand\orcidicon}}
}
			
\begin{document}

\title{Deep Cross-Modality and Resolution Graph Integration for Universal Brain Connectivity Mapping and Augmentation}

\titlerunning{Deep Cross-Modality and Resolution Graph Integration}  

\author{Ece Cinar$^\dagger$\index{Cinar, Ece}, Sinem Elif Haseki$^\dagger$\index{Haseki, Sinem Elif}, Alaa Bessadok\index{Bessadok, Alaa} \and Islem Rekik\orcidA{} \index{Rekik, Islem}\thanks{ {corresponding author: irekik@itu.edu.tr, \url{http://basira-lab.com}.  $^\dagger$: co-first authors. }}}

\institute{BASIRA Lab, Faculty of Computer and Informatics Engineering, Istanbul Technical University, Istanbul, Turkey (\url{http://basira-lab.com/})}

\authorrunning{E. Cinar et al.}  

\maketitle              

\begin{abstract}
The connectional brain template (CBT) captures the shared traits across all individuals of a given population of brain connectomes, thereby acting as a fingerprint. Estimating a CBT from a population where brain graphs are derived from diverse neuroimaging modalities (e.g., functional and structural) and at different resolutions (i.e., number of nodes) remains a formidable challenge to solve. Such network integration task allows for learning a rich and universal representation of the brain connectivity across varying modalities and resolutions. The resulting CBT can be substantially used to generate entirely new multimodal brain connectomes, which can boost the learning of the downs-stream tasks such as brain state classification. Here, we propose the Multimodal Multiresolution Brain Graph Integrator Network (i.e., M2GraphIntegrator), \emph{the first multimodal multiresolution graph integration framework that maps a given connectomic population into a well-centered CBT.} M2GraphIntegrator first unifies brain graph resolutions by utilizing resolution-specific graph autoencoders. Next, it integrates the resulting fixed-size brain graphs into a universal CBT lying at the center of its population. To preserve the population diversity, we further design a novel clustering-based training sample selection strategy which leverages the most heterogeneous training samples. To ensure the biological soundness of the learned CBT, we propose a topological loss that minimizes the topological gap between the ground-truth brain graphs and the learned CBT. Our experiments show that from a single CBT, one can generate realistic connectomic datasets including brain graphs of varying resolutions and modalities. We further demonstrate that our framework significantly outperforms benchmarks in reconstruction quality, augmentation task, centeredness and topological soundness.

\end{abstract}


\section{Introduction}


Modern network science opens new frontiers of representing the complex functionality and structure of biological systems by analyzing the intercommunication within their fundamentals \cite{ideker2001new}. The wealth of technological advances in the field of neuroscience paves the way for gathering massive and high-quality biological datasets such as Human Connectome Project \cite{HCP}, Southwest University Longitudinal Imaging Multimodal (SLIM) Brain Data Repository \cite{qiusouthwest} and UK Biobank \cite{biobank2014uk} using different magnetic resonance imaging (MRI) modalities including functional, structural T1-weighted and diffusion MRI. Representing such connectomic datasets using graphs (i.e., networks) aims to reveal the complex interconnections between brain regions. More specifically, each brain graph allows to investigate particular connectivity patterns and functionalities of the brain elements, where each anatomical brain region of interest (i.e., ROI) is represented with a node and the biological connectivity between two ROIs is represented by weighted edges \cite{fornito2015connectomics,fornito:2016,van2019cross}. Using graphs that are derived from such rich multimodal datasets serves as an exemplary tool for examining the human brain structure and state \cite{seidlitz2018morphometric,holmes2015brain} by mapping the brain wiring at the \emph{individual} level. 

In addition to fingerprinting the brain of an individual, graph representations allow for mapping brain connectivity at the \emph{population} level, thereby distinguishing between contrasting states (e.g., healthy versus unhealthy) of different populations. Emerging studies focused on learning how to integrate a set of unimodal single-resolution brain graphs into a single connectome (i.e., connectional brain template) that encodes the shared traits across the individuals of the population \cite{rekik2017estimation,dhifallah2020estimation,gurbuz2020deep}.
Despite their overwhelming success, existing methods \cite{alaa} are not particularly designed to handle \emph{multimodal multiresolution} connectomic datasets, which, if solved, can pave the way for holistically detecting anomalies and abnormalities across varying brain networks. Specifically, generating a universal connectional brain template (i.e., CBT) from a \emph{multimodal multiresolution connectomic population} remains an uncharted territory \cite{alaa}. By mitigating such a challenging issue, a single compact representation, from which one can span new multimodal multiresolution brain graph populations for data augmentation \cite{nalepa2019data,perl2020data}, can be learned to reveal typical and atypical alterations in the brain connectome across modalities and various individuals. One can also leverage the universal CBT for \emph{graph augmention} to alleviate clinical data scarcity \cite{sserwadda2021topology,khan2020post,you2020graph,nalepa2019data} in classification and regression tasks \cite{perez2017effectiveness,mikolajczyk2018data,GRAA2019108344,wong2016understanding,du2018classification}.

\textbf{Related work.} Existing works tailored for graph integration or fusion in general are limited to training on unimodal, single-resolution brain networks \cite{wang2014similarity,rekik2017estimation,dhifallah2020estimation,gurbuz2020deep}. For example, based on message passing between the neighbors of a particular node, similarity network fusion (SNF) \cite{wang2014similarity}  learns how to integrate a set of biological graphs by diffusing the local connectivity of each individual graph across the global connectivity of all samples in the population in an iterative manner. Still, such method cannot handle multiresolution graphs as well as heterogeneous samples drawn from multimodal distributions. Later on, \cite{rekik2017estimation} proposed a novel method for estimating a CBT (also termed with brain network atlas) over a population of brain networks which are derived from the same modality by exploiting diffusive-shrinking and fusing graph techniques. However, the mathematical formalization of the proposed graph diffusion and fusion method is not capable of handling multigraph population, where each sample is represented by a set of graphs. 
To remedy the lack of methods for \emph{multigraph data integration}, where a multigraph allows for multiple edges connecting two nodes, \cite{dhifallah2020estimation} introduced a novel approach for multi-view graph construction. 
However, such method utilizes disparate learning modules that learn independently without any feedback mechanism between them; as such the errors accumulate throughout the dichotomized learning pipeline. To address this issue, \cite{gurbuz2020deep} introduced Deep Graph Normalizer (DGN) framework, the first graph neural network that integrates a population of fixed-size multigraphs in an end-to-end learnable way. Although compelling, DGN is limited to aggregating the information only across multi-view brain graphs with a \emph{fixed resolution}. Besides, it  relies on a random sampling technique to generate CBTs, which is agnostic to \emph{data heterogeneity}. Moreover, DGN uses edge-conditioned convolution, which is not fundamentally tailored for easing the memory consumption, thereby undermining the population representative CBT estimation for large-scale graph populations. Other related works \cite{demir2020clustering,sauglam2020multi} focused only on integrating single-resolution brain network populations for disorder profiling and CBT learning. We note a few works that were also dedicated to brain graph super-resolution \cite{isallari2020gsr,mhiri2021stairwaygraphnet,mhiri2020brain}, which primarily aimed to generate brain graphs across different resolutions rather then integrating them. 

To address all these limitations, we propose \emph{Multimodal Multiresolution Graph Integrator} (M2GraphIntegrator) Network, the first framework for integrating a population of \emph{multimodal multi-resolution} brain networks into a centered and representative CBT. Tapping into the nascent field of GNNs, we design a set of resolution-specific autoencoders to map a given population of brain networks of different resolutions derived from multiple modalities to a shared embedding space. Next, given the learned embeddings, we generate the CBT through the integrator network, which is an architecture specialized in learnable embedding integration. To train our framework, we design a novel CBT \emph{ centeredness loss} that ensures the heterogeneity of training samples via clustering. As such, the selected training samples from different clusters represent each and every distribution present in the input graph population. In that way, our estimated CBT can capture the connectivity patterns across all subjects in a diverse population. Furthermore, to preserve the brain graph topology in the integration process, we propose a novel \emph{topology loss} which aims to minimize the topological gap between the ground-truth and the reconstructed brain graphs in terms of node strength, a measure quantifying the local hubness of each brain node (i.e., anatomical region of interest).

\begin{figure}[ht!] 
\includegraphics[width=\textwidth]{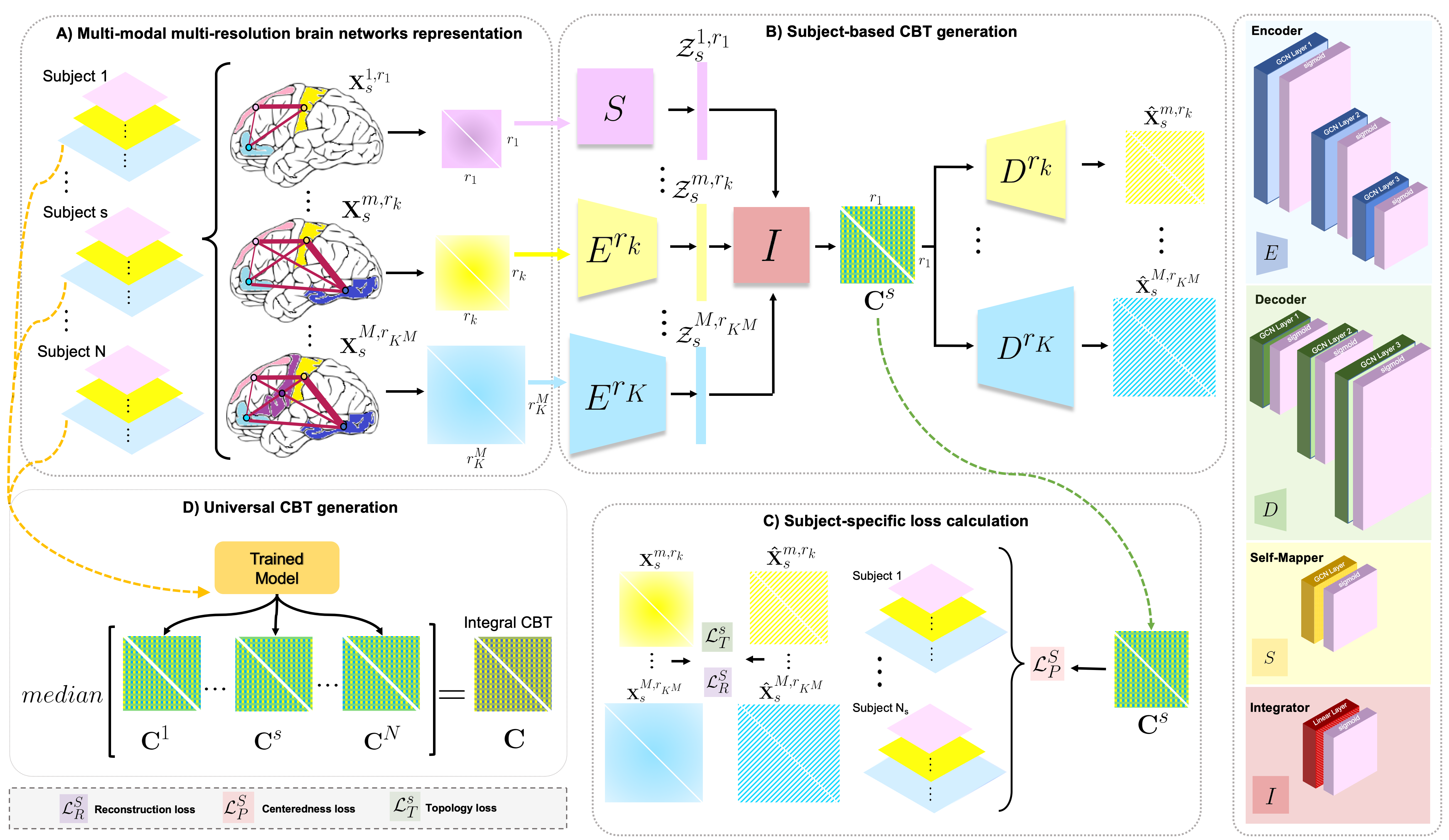}
\caption{\emph{Overview of the proposed Multi-modal Multi-resolution Graph Integrator (M2GraphIntegrator) architecture for estimating a centered connectional brain template from a given population.} \textbf{(A) Multi-modal multi-resolution brain network representation.} We represent each subject in the population by multiple connectivity matrices, each denoted by $\mathbf{X}_s^{m,r_k}$ $\in  \mathbb{R}^{r_k \times r_k}$. \textbf{(B) Subject-based CBT generation.} Our framework consists of 3 co-learning modules: the resolution-specific graph autoencoders, the self-mapper and the integrator. We generate the subject-based CBT by integrating the feature vector embeddings $\mathcal{Z}_s^{m,r_k}$ of each encoder and the self-mapper. \textbf{(C) Subject-specific loss calculation.} For each subject $s$, we calculate both reconstruction and topological losses using the whole training set. As for the CBT centeredness loss, using clustering we select a heterogeneous subset of training samples against which we evaluate the centeredness of the learned subject-specific CBT. \textbf{(D) Universal CBT generation}. To capture the most centered connectional patterns across all subjects, we feed each brain multigraph through our trained model to generate subject-based CBTs. Next, we perform element-wise median operation to estimate the \emph{universal CBT}. To simplify the illustration, we denote the encoder $E^{r_K^M}$ by $E^{r_K}$ and $D^{r_K^M}$ by $D^{r_K}$.}
\label{fig:2}
\end{figure}

\section{Proposed Method}

In this section, we present the main steps of our CBT estimation framework from multimodal multi-resolution brain networks. \textbf{Fig.}~\ref{fig:2} provides an overview of the key three steps of the proposed framework: \textbf{A)} representation of multimodal and multi-resolution brain networks in a population, \textbf{B)} generation of subject-based CBT, \textbf{C)} subject-specific loss calculation, and \textbf{D)} estimation of the universal CBT.

\textbf{A- Multi-modal multi-resolution brain networks representation}. 
Given a connectomic population, each subject is represented by multiple brain networks of different resolutions derived from different neuroimaging modalities such as functional and structural MRI  (\textbf{Fig.}~\ref{fig:2}\textbf{-A}). Such brain networks do not necessarily belong to the same resolution set (i.e., they might have different number of nodes thus different number of edges). Therefore, we represent each subject $s$ in the population as follows:

\centerline{$ \mathcal{X}_s = \{ {X}_s^{m} \}_{m=1}^{M},\hspace{0.2cm} {X}_s^m = \{ \mathbf{X}_s^{m,r_k} \}_{k=1}^{K^m} \text{, where } r_1 < \dots < r_k < \dots < r_{K^m}  $}

${X}_s^{m}$ denotes the modality-specific brain networks set derived from modality $m$ and $\mathcal{X}_s$ stands for the overall set encapsulating each and every modality-specific brain networks set of subject $s$. $\mathbf{X}_s^{m,r_k} \in  \mathbb{R}^{r_k \times r_k}$ represents a connectivity matrix (i.e. adjacency matrix) of resolution $r_k$ belonging to ${X}_s^m$. We represent the total number of resolutions derived from modality $m$ by $K^m$ and further denote by $K$ the total number of resolutions across all modalities. Since each connectivity matrix is symmetric, we vectorize it into a feature vector by taking the elements in its lower triangular part. Specifically, for each subject, we represent each brain graph with a feature vector $\mathbf{V}_s^{m,r_k}$ $\in$ $\mathbb{R}^{1 \times r'_k}$, where $r'_k$ = $r_k \times (r_k - 1) \over 2$. We note in what follows that our GCN is trained in a subject-based fashion where each subject is represented by a single-node brain graph.

\textbf{B- Subject-based CBT generation}. To estimate a subject-based CBT $\mathbf{C}^s$, we design 3 GCN-based modules that co-learn during the training process: the \emph{resolution-specific graph autoencoders}, the \emph{self-mapper} and the \emph{integrator} (\textbf{Fig.}~\ref{fig:2}\textbf{-B}).

\emph{Resolution-specific graph encoder.}
In order to generate a subject-based CBT, we first propose to reduce the resolution of differently scaled brain networks. To do so, we introduce a set of resolution-specific graph encoders $\{E^{r_k}\}_{k=2}^K$, where each $E^{r_k}$ maps the feature vectors of a network at resolution $r_k$ into a shared lower embedding space at the lowest existing graph resolution $r_1$. Our encoders learn to capture the shared-traits across multi-resolution brain networks. We build the encoders by stacking three GCN blocks each including a GCN layer followed by sigmoid non-linearity and a dropout function. Each GCN layer performs the graph convolution operation \cite{kipf2017semisupervised} defined as follows: $\mathbf{V}^l = \mathbf{\hat{D}}^{-1/2} \mathbf{\hat{A}} \mathbf{\hat{D}}^{-1/2} \mathbf{V}^{l-1} \mathbf{\Theta}^l$, where $\mathbf{V}^l$ denotes the feature vector embedding at layer $l$, $\mathbf{\hat{D}}$ denotes the diagonal degree matrix, $\mathbf{\hat{A}}$ denotes the adjacency matrix including self-connectivities and $\mathbf{\Theta}^l$ denotes the learnable parameter for each layer $l$. For simplicity, we choose $\mathbf{V}$ as a representation of $\mathbf{V}_s^{m,r_{k}}$ which is the feature vector of a subject $s$ with a resolution $r_k$ and modality $m$. We note that feature vector embedding of the last layer (i.e., third layer of $E^{r_k}$) is denoted by $\mathcal{Z}$ where $\mathcal{Z}$ = $\mathbf{V}^L$ (i.e., feature vector embedding at layer $L$, where $L=3$ in our case).

As the first step of estimating $\mathbf{C}^s$, we pass the feature vectors of subject $s$ each denoted by $\mathbf{V}_s^{m,r_k}$ through their corresponding resolution-specific graph encoders in order to map them to the shared embedding space of size $r'_1$. We represent the low-dimensional embedding of each feature vector by $\mathcal{Z}_s^{m,r_k}$ $\in$ $\mathbb{R}^{1 \times r'_1}$. We note that the minimal resolution brain graphs of subject $s$ are passed to another architecture called the \emph{self-mapper}, which we will detail in the following section.

\emph{Self-mapper.} The self-mapper is an architecture aiming to generate feature vector embeddings which capture the shared-traits across subjects of a given population. It maps minimal-resolution feature vectors into the embedding space shared among the resolution-specific graph encoders. Since the self-mapper does not alter the resolution of its input feature vectors, it cannot be identified as an encoder. However, the self-mapper and resolution-specific graph encoders resemble each other in terms of generating population-representative feature vector embeddings. The self-mapper consists of a single GCN block built by stacking a GCN layer, sigmoid non-linearity and dropout function and is denoted by $S$ throughout our framework (\textbf{Fig.}~\ref{fig:2}-B). In line with the purpose of estimating $\mathbf{C}^s$, we pass each minimal-resolution feature vector $\mathbf{V}_s^{m,r_{1}}$ of subject $s$ through the self-mapper and denote their embeddings by $\mathcal{Z}_s^{m,r_1}$.

\emph{Integrator.} We introduce an integrator module $I$ to integrate the feature embeddings $\mathcal{Z}_s^{m,r_k}$ generated by the different resolution-specific graph encoders $\{E^{r_k}\}_{k=2}^K$ and the self-mapper $S$ into a single representation --i.e., the \emph{subject-based} CBT. It mainly encapsulates multiple integration blocks, each composed of a linear layer followed by sigmoid non-linearity. Mainly, to estimate $\mathbf{C}^s$ we first pass feature embeddings of subject $s$ through their corresponding integration blocks. Second, we average the integration block outputs and generate the subject-based CBT $\mathbf{C}^s$ in its vectorized version (\textbf{Fig.}~\ref{fig:2}-B). Finally, we derive the matrix representation of $\mathbf{C}^s$ by simple antivectorization. 


\emph{Resolution-specific graph decoder.} 
We design our M2GraphIntegrator framework in a way that each resolution-specific graph encoder $E^{r_k}$ has a corresponding resolution-specific graph decoder denoted by $D^{r_k}$. These encoder-decoder pairs are symmetric since both architectures consist of identical graph convolutional blocks in a reversed order. Even though decoders are not directly involved in the subject-based CBT estimation process, they play a substantial role in the overall framework by forcing the encoders, the self-mapper and the integrator to better learn the population graph representation. To achieve this, we propose two losses: the \emph{reconstruction loss} and the \emph{topology loss} which we will address in the following sections. Specifically, each decoder $D^{r_k}$ maps the learned $\mathbf{C}^s$ into a higher embedding space of size $r_k$ in order to reconstruct the initial feature vector of subject $s$, which is antivectorized into the reconstructed brain connectivity matrix $\mathbf{\hat{X}}_s^{m,r_k}$. We note that our resolution-specific graph decoders can be further leveraged for \textbf{multimodal brain network data augmentation} by perturbing the learned population CBT.

\textbf{C- Subject-specific loss calculation}. Once we generate $\mathbf{C}^s$ from the integrator, we calculate a \emph{centeredness loss} inspired by the subject normalization loss (SNL) proposed in \cite{gurbuz2020deep}. However, SNL cannot preserve the data heterogeneity of the population which might result in a limited representation that fails to fully capture the spectrum of brain connectivity variability across subjects. To solve this problem, we propose a different method for selecting a subset $\mathcal{D}_S$ of our training dataset $\mathcal{D}$. Specifically, we employ a clustering-based sampling method (e.g., K-means or hierarchical clustering) rather than solely using random sampling. For \emph{centeredness loss} calculation, we first vectorize the input connectivity matrices and stack the resulting feature vector embeddings. Next, to select a $\mathcal{D}_S$, we employ a clustering step to sample subjects from different clusters and produce their embedding vectors $\mathcal{Z}_s^{m,r_k}$ using the encoder $E^{r_k}$. Finally, we obtain $\mathbf{Z}_s^{m,r_k}$ by antivectorizing $\mathcal{Z}_s^{m,r_k}$ for each sample in $\mathcal{D}_S$ and calculate the mean Frobenius distance (MFD) with respect to $\mathbf{C}^s$: $\mathcal{L}_{P}^s =  {\sum_{m=1}^{M}\sum_{k=1}^{K^m} \sum_{n=1}^{N_S}\lvert\lvert {\mathbf{C}^s - \mathbf{Z}_s^{m, r_k}}\lvert\lvert_2^2}$.
To ensure that the decoded network preserves the initial traits present in the ground-truth graphs, we introduce a \emph{reconstruction loss} which computes the MFD between the ground-truth connectivity matrices $\mathbf{X}_s^{m,r_k}$ and their reconstructed matrices $\mathbf{\hat{X}}_s^{m,r_k}$ for each subject $s$. We define it as follows: $\mathcal{L}_{R}^s = {\sum_{m=1}^{M}\sum_{k=1}^{K^m} \lvert\lvert {\mathbf{X}_s^{m,r_k} - \mathbf{\hat{X}}_s^{m, r_k}}\lvert\lvert_2^2}$.
We further propose a new \emph{topology loss} to enforce the connectivity strength of the brain regions in the reconstructed brain graphs $\mathbf{\hat{X}}_s^{m,r_k}$ to be similar to those of the ground-truth networks $\mathbf{X}_s^{m,r_k}$. More specifically, we generate for each subject in the population $\mathbf{\hat{X}}_s^{m,r_k}$ by passing the vectorized $\mathbf{C}^s$ through the decoder ${D}^{r_k}$. Next, we calculate the node strength vectors $\mathbf{P}_s^{m,r_k}$ and $\mathbf{\hat{P}}_s^{m,r_k}$ by summing up then normalizing over the rows of $\mathbf{X}_s^{m,r_k}$ and $\mathbf{\hat{X}}_s^{m,r_k}$, respectively. Hence, we define it as follows: $\mathcal{L}_{T}^s =  {\sum_{m=1}^{M}\sum_{k=1}^{K^m} \lvert\lvert {\mathbf{P}_s^{m,r_k} - \mathbf{\hat{P}}_s^{m, r_k}}\lvert\lvert_1}$.
By combining the three sub-losses (\textbf{Fig.}~\ref{fig:2}\textbf{-C}), we define the total subject-specific loss for a training subject $s$ as follows:

\begin{equation*}
\small
\mathcal{L}_{J}^s =  {\sum_{m=1}^{M}\sum_{k=1}^{K^m}\Big( \:\underbrace{\lvert\lvert {\mathbf{X}_s^{m,r_k} - \mathbf{\hat{X}}_s^{m, r_k}}\lvert\lvert_2^2}_\text{Reconstruction loss}  + \lambda_1\underbrace{ \lvert\lvert {\mathbf{P}_s^{m,r_k} - \mathbf{\hat{P}}_s^{m, r_k}}\lvert\lvert_1}_\text{Topology loss} + \lambda_2\underbrace{\textstyle  \sum_{n=1}^{N_S}{\lvert\lvert\mathbf{C}^s -\mathbf{Z}_{n}^{m, r_k}\lvert\lvert_2^2}}_\text{Centeredness loss}\: \Big)}
\end{equation*}

In that way, our proposed loss not only captures the topological structure and information of different networks, but also the shared traits across subjects of the population.

\textbf{D- Universal CBT generation}. Since our ultimate goal is to generate a population representative CBT rather than a subject-based CBT, we further propose an additional step in our framework. Since each $\mathbf{C}^s$ generated in the previous step is biased by a particular subject, we need to acquire a centered CBT that represents the heterogeneous population. To mitigate this issue, we perform element-wise median operation on all generated subject-based CBTs (i.e., $\mathbf{C}^s$) as follows: $\mathbf{C} = {median}[\mathbf{C}^1,\mathbf{C}^2,...,\mathbf{C}^N]$, where $N$ represents the number of training subjects (\textbf{Fig.}~\ref{fig:2}\textbf{-D}). As a result, we estimate $\mathbf{C}$ the integral CBT that represents each and every subject in a multimodal multi-resolution brain graph population. 

\section{Results and Discussion}
\textbf{Connectomic dataset and hyperparameter setting.}
We trained and tested our framework on a connectomic dataset derived from the Southwest University Longitudinal Imaging Multimodal (SLIM) Brain Data Repository \cite{qiusouthwest}. The dataset consists of 279 young healthy subjects, each represented by two brain networks of resolutions (i.e., ROIs) 35 and 160 derived from T1-weighted  (morphological network) and resting-state functional MRI (functional network). We benchmarked our \textbf{M2GraphIntegrator} including the topological loss \textbf{(T)} with its two \textbf{(K)} K-means  and \textbf{(H)} hierarchical clustering variants against four ablated versions: \textbf{Ablated (K)} and \textbf{Ablated (H)} employ K-means and hierarchical clustering without integrating the topology loss while \textbf{Ablated (R+T)} and \textbf{Ablated (R)} employ random sampling with and without including the topology loss, respectively. We initialized two clusters for each clustering-based sampling method and selected 10 training subjects at each epoch. Prior to calculating the population \emph{centeredness loss}, we provided an extra training of 100 epochs for the resolution-specific graph autoencoder architecture to achieve more reliable and steady results in the graph reconstruction block. We used grid search to tune the hyperparameters $\lambda_1$ and $\lambda_2$ of our loss function $\mathcal{L}_{J}^s$ and set them to $2$ and $0.5$, respectively. We used Adam optimizer and set the learning rate to $0.0001$.

\begingroup
\renewcommand{\arraystretch}{1.3}
\begin{table}[htp!]
\centering
\begin{tabular}{c|cccc}
\hline\hline
 \multirow{2}{*}{Model Variation} & \multicolumn{4}{c}{Evaluation Measure}  \\ 
\cline{2-5} 
        &  \colorbox{pink!40}{Centeredness}   &  \colorbox{blue!20}{Top. Soundness} &  \colorbox{green!30}{KL Divergence} &  \colorbox{yellow!40}{Pairwise Dist.}  \\   \cline{1-5}
  Ablated (R)    &  19.0727   & 5.9825 & 0.9764 & 0.1495000 \\ 
  Ablated (K)   &  19.0702   & 5.9767 & 0.9766  &  0.1495006\\ 
  Ablated (H)    &  19.0755  & 5.9831& 0.9765 & 0.1495009\\ 
 Ablated (R+T)    &  17.9822  & 5.8448 &  0.7245 &  0.1421824 \\ 
M2GraphIntegrator (K+T)    & $\underline{17.9819}$  & $\underline{5.8439}$ & $\mathbf{0.7241}$ &  $\underline{0.1421812}$ \\  
 M2GraphIntegrator (H+T)   &  $\mathbf{17.9783}$  & $\mathbf{5.8434}$ &  $\underline{0.7243}$ & $\mathbf{ 0.142180}$\\ 
 \hline\hline
\end{tabular}
    \caption{\emph{CBT evaluation results using different measures.} Centeredness and topological soundness evaluate the quality of the generated CBT. KL divergence and pairwise distance evaluate the ability of the learned CBT in generating sound multimodal brain networks at different resolutions for data augmentation. For each metric, we highlight in bold the best performing method and underline the second best.  Both M2GraphIntegrator (K+T) and (H+T) significantly outperformed ablated comparison methods ($p-value<0.01$ using two-tailed paired t-test). \textbf{K}: K-means clustering. \textbf{H}: hierarchical clustering. \textbf{R}: random sampling. \textbf{T}: topological loss.}
\label{tab:3}
\end{table}
\endgroup

\textbf{Evaluation of universal CBT centeredness and topological soundness.} A representative CBT lies at the center of its population, hence it needs to achieve the minimal distance to all subjects in the population. We first use 5-fold cross-validation to split the data into training and testing folds where the CBT is learned from the training set and evaluated against the unseen test set. To evaluate the centeredness of the universal CBT, we first encode the testing functional graphs with $160\times160$ resolution by the trained $E^{160}$. Next, we compute the mean Frobenius distance between the CBT matrix learned from the training set and each morphological and encoded functional matrix of testing subjects. Next, to assess the topological soundness of the estimated CBT, we compute the Euclidean distance between their corresponding node strength representations. Lower values of the centeredness and topological soundness measures demonstrate that the generated CBT is more representative and topology-aware. \textbf{Table}~\ref{tab:3} shows the significant outperformance of our M2GraphIntegrator across all evaluation measures using both K-means and hierarchical clustering methods. Notably, these results show that our model learned using the proposed topology loss function significantly outperforms ($p < 0.01$) the comparison methods in preserving the topological properties of the ground-truth networks.

\textbf{Evaluation of multimodal network augmentation from the learned universal CBT.}  Assuming that the \emph{universal CBT} spans all domains across modalities and resolutions, it can be easily utilized to generate new brain networks for potential downstream learning tasks (e.g., connectome regression \cite{shen2017using}). First, we simulate 279 random networks of the same CBT size and distribution. Next, we regularize each random network by averaging it with the universal CBT. Next, we feed each average network to our resolution-specific graph decoders to generate multimodal networks. To assess the realness of the generated networks, we compute the Kullback-Leibler divergence \cite{kullback1997information} between the ground-truth and the augmented brain networks as well as their average pairwise Euclidean distance (\textbf{Table}~\ref{tab:3}). Hence, a lower result in both metrics represents higher similarity between the ground-truth and the networks generated from our \emph{universal} CBT (\textbf{Table}~\ref{tab:3}). Remarkably, the universal CBT by our methods \textbf{M2GraphIntegrator (K+T)} and \textbf{M2GraphIntegrator (H+T)} generates more significantly ($p<0.01$) realistic multimodal brain graphs compared to the benchmarks.

\section{Conclusion}
In this paper, we proposed Multi-modal Multi-resolution Graph Integrator which is the first graph neural network framework that estimates a connectome population fingerprint given \emph{multimodal multi-resolution} brain networks. Our method has three compelling strengths: (i) the autoencoder learning task with joint multi-resolution GCN-based autoencoders, facilitating its customizability to any graph resolution, (ii) the design of the \emph{clustering-based} training sampling in the centeredness loss computation to learn a well-representative CBT of the population heterogeneity and (iii) the proposal of the \emph{topology loss} to estimate a topologically sound CBT. Our estimated CBTs will not only pave the way for easier brain disorder diagnosis by revealing deviations from the healthy population but also remedy data scarcity by augmenting new brain networks. In our future work, we will use our model to learn universal CBTs of various healthy and disordered brain connectivity datasets including functional, morphological, and structural connectomes. Besides, we will refine our architecture by integrating a novel graph new edge-convolution that operates on large-scale graphs without memory overloading. 

\section{Acknowledgements}

This work was funded by generous grants from the European H2020 Marie Sklodowska-Curie action (grant no. 101003403, \url{http://basira-lab.com/normnets/}) to I.R. and the Scientific and Technological Research Council of Turkey to I.R. under the TUBITAK 2232 Fellowship for Outstanding Researchers (no. 118C288, \url{http://basira-lab.com/reprime/}). However, all scientific contributions made in this project are owned and approved solely by the authors.

\newpage
\bibliography{Biblio3}
\bibliographystyle{splncs}
\end{document}